\documentclass[aps,prl,twocolumn,superscriptaddress,preprintnumbers,%
               showpacs,nofootinbib]{revtex4}
\newcommand{\PRE}[1]{}       

\usepackage{bm}
\usepackage{epsfig}

\newcommand{\postscript}[2]{\setlength{\epsfxsize}{#2\hsize}
   \centerline{\epsfbox{#1}}}

\def\eslt{\not\!\!\!{E_T}}
\def\to{\rightarrow}

\def\bi{\begin{itemize}}
\def\ei{\end{itemize}}

\def\tg{\tilde g}

\def\tell{\tilde\ell}
\def\tq{\tilde q}
\def\tw{\widetilde W}
\def\tz{\widetilde Z}
\def\alt{\stackrel{<}{\sim}}
\def\agt{\stackrel{>}{\sim}}
\def\be{\begin{equation}}  
\def\ee{\end{equation}}

\newcommand\plb[3]{{\it Phys\ Lett.\ }{\bf B #1} (#2) #3}
\newcommand\jhep[3]{{\it J.\ High Energy Phys.\ }{\bf #1} (#2) #3}
\newcommand\npb[3]{{\it Nucl.\ Phys.\ }{\bf B #1} (#2) #3}
\newcommand\ijmpa[3]{{\it Int. \ J.\ Mod.\ Phys.\ }{\bf A #1} (#2) #3}
\newcommand\epjc[3]{{\it Eur. \ Phys.\ J. }{\bf C #1} (#2) #3}
\newcommand\jphg[3]{{\it J.\ Phys.\ }{\bf G #1} (#2) #3}
\newcommand\zpc[3]{{\it Z.\ Physik\ }{\bf C #1} (#2) #3}
\newcommand\prD[3]{{\it Phys.\ Rev.\ }{\bf D #1} (#2) #3}
\begin{document}

\preprint{UH-511-1186-12}

\title{
\PRE{\vspace*{1.5in}}
$Wh$ plus missing-$E_T$ signature \\
from gaugino pair production at the LHC
\PRE{\vspace*{0.3in}}
}

\author{Howard Baer}
\affiliation{Dept. of Physics and Astronomy,
University of Oklahoma, Norman, OK, 73019, USA
\PRE{\vspace*{.1in}}
}
\author{Vernon Barger}
\affiliation{Dept. of Physics,
University of Wisconsin, Madison, WI 53706, USA
\PRE{\vspace*{.1in}}
}
\author{Andre Lessa}
\affiliation{Instituto de F\'{i}sica, 
Universidade de S\~{a}o Paulo, S\~{a}o Paulo - SP, Brazil
\PRE{\vspace*{.1in}}
}
\author{Warintorn Sreethawong}
\affiliation{Dept. of Physics and Astronomy,
University of Oklahoma, Norman, OK, 73019, USA
\PRE{\vspace*{.1in}}
}
\author{Xerxes Tata}
\affiliation{Dept. of Physics and Astronomy,
University of Hawaii, Honolulu, HI 96822, USA
\PRE{\vspace*{.1in}}
}


\begin{abstract}
\PRE{\vspace*{.1in}} In SUSY models with heavy squarks and gaugino mass
unification, the gaugino pair production reaction $pp\to\tw_1^\pm\tz_2$
dominates gluino pair production for $m_{\tg}\agt 1$ TeV at LHC with
$\sqrt{s}=14$ TeV (LHC14).  For this mass range, the two-body decays
$\tw_1\to W\tz_1$ and $\tz_2\to h\tz_1$ are expected to dominate the
chargino and neutralino branching fractions.  By searching for $\ell
b\bar{b}+\eslt$ events from $\tw_1^\pm\tz_2$ production, we show that
LHC14 with 100 fb$^{-1}$ of integrated luminosity becomes sensitive to
chargino masses in the range $m_{\tw_1}\sim 450-550$ GeV corresponding
to $m_{\tg}\sim 1.5-2$ TeV in models with gaugino mass unification.  For
$10^3$ fb$^{-1}$, LHC14 is sensitive to the $Wh$ channel for
$m_{\tw_1}\sim 300-800$ GeV, corresponding to $m_{\tg}\sim 1-2.8$ TeV,
which is comparable to the reach for gluino pair production followed by
cascade decays. 
The $Wh+\eslt$ search channel opens up a new complementary avenue for SUSY
searches at LHC, and serves to point to SUSY as the origin of any new
physics discovered via multijet and multilepton + $\eslt$ channels

\end{abstract}

\pacs{12.60.-i, 95.35.+d, 14.80.Ly, 11.30.Pb}

\maketitle

One of the major goals of the CERN LHC is to discover or rule out as best as possible 
particle physics theories based on weak scale supersymmetry\cite{wss} (SUSY). 
Recent SUSY searches by ATLAS\cite{atlas} and CMS\cite{cms} using $pp$ collisions 
at $\sqrt{s}=7$~TeV (LHC7) have been performed in the context of the 
minimal supergravity (mSUGRA or CMSSM) model\cite{msugra}. 
In this model, all scalar particles receive a common mass $m_0$ and all gauginos acquire a common 
mass $m_{1/2}$ at the grand-unified scale $M_{GUT}\sim 2\times 10^{16}$ GeV. 
Assuming the MSSM as the low energy effective theory, 
the various soft SUSY breaking parameters are then evolved via renormalization group equations to the weak scale, 
whereupon the various sparticle masses and mixings can be calculated.

Based on non-observation of signal
events at rates expected beyond Standard Model backgrounds in $\sim 1$ fb$^{-1}$ of data, 
ATLAS and CMS have been able to plot excluded regions in the $m_0\ vs.\ m_{1/2}$ plane of the 
mSUGRA model. 
These exclusion limits correspond to $m_{\tg}\agt 1$ TeV in the case where $m_{\tg}\sim m_{\tq}$, 
and $m_{\tg}\agt 600$ GeV in the case where $m_{\tq}\gg m_{\tg}$ 
(the case with $m_{\tq}\ll m_{\tg}$ doesn't occur in the mSUGRA model).

At the present time, ATLAS and CMS have each accumulated more than 5
fb$^{-1}$ of data, and analyses of this data set are anxiously awaited
by the particle physics community.  Further running in 2012 is expected
to net 10-30 fb$^{-1}$ of integrated luminosity at $\sqrt{s}=7$ TeV.  It
is expected that LHC will be shut down during the year 2013 for
upgrading, and running will resume in 2014  at a center-of-mass energy
		     close to the LHC design value,
$\sqrt{s}\sim 14$ TeV (LHC14).

In evaluating the reach of LHC for SUSY particles, searches tend to
focus on gluino pair production ($\tg\tg$), squark pair production
($\tq\tq$) and gluino-squark production ($\tg\tq$), since strongly
interacting sparticles are expected to be produced at the larger rates
than chargino/neutralino or slepton pair production\cite{bcpt1}.  Since
the gluinos and squarks are typically amongst the most massive members
of the entire SUSY particle spectrum, they are expected to {\it cascade
decay}\cite{cascade} via lengthy chains into final states containing
numerous jets, isolated leptons and missing transverse energy $\eslt$.

To estimate the SUSY reach of any collider, first the SUSY particle
masses and mixings must be calculated for a given model.  Then, the
various sparticle pair production reactions must be generated according
to their relative probabilities (cross sections), and unstable
sparticles  allowed to decay using the calculated decay widths and
branching fractions. Incorporation of initial and final state QCD
radiation, hadronization of partons, further decays of unstable
particles and a modeling of the underlying collider event will then
allow for a hopefully realistic determination of what sparticle
pair production events look like at the LHC.

The reach of LHC14 for 10 fb$^{-1}$ was first evaluated in
Ref. \cite{bcpt1} for events with multi-jets $+\eslt$, and later in
Ref.~\cite{bcpt2} for events containing various isolated leptons plus
jets $+\eslt$ topologies. Updated projections for 100 fb$^{-1}$ were
plotted in Ref. \cite{bbbkt}, where it was found that the LHC14 reach
can extend to $m_{\tg}\sim 3$ TeV for $m_{\tq}\sim m_{\tg}$, while the
reach is to $m_{\tg}\sim 1700$ GeV for $m_{\tq}\gg m_{\tg}$.  The LHC7
reach was shown in Ref. \cite{lhc7} for integrated luminosities up to 2
fb$^{-1}$ and later 30 fb$^{-1}$, while the reach for LHC14 (and LHC10)
was calculated in Ref. \cite{lhc10} for integrated luminosities up to
1000-3000 fb$^{-1}$.\footnote{In Ref. \cite{lhc10}, the 100 fb$^{-1}$
reach of LHC14 was found to extend to $m_{\tg}\sim 2.1$ TeV for $m_0\sim
3$~TeV, for $\tan\beta =45$.  In this region, squarks have not
completely decoupled in the focus point region, so the reach is somewhat
higher than expected for the squark decoupling regime ($m_{\tq}\agt
5$~TeV at LHC14).}  In all these studies, work was performed in the
$R$-parity conserving mSUGRA model with the lightest neutralino $\tz_1$
as lightest SUSY particle (LSP).\footnote{ For LHC reach in GMSB, see
Ref. \cite{lhc_gmsb}; for reach in AMSB, see Ref. \cite{lhc_amsb} and
for reach in inoMSB, see Ref. \cite{lhc_ino}; for reach in mSUGRA with
hadronic RPV neutralino decays, see \cite{lhc_rpv}.  } A stable
neutralino LSP provides a distinctive $\eslt$ signature at LHC, and may
be associated with a dark matter WIMP.

In models with gaugino mass unification ({\it i.e.} the soft SUSY breaking gaugino masses
$M_1$, $M_2$ and $M_3$ unify to a common value $m_{1/2}$ at energy scale $Q=M_{GUT}$), the 
{\it weak scale} gaugino masses are expected to be (aside from 2-loop RG effects) 
in the ratio $M_1:M_2:M_3\sim 1:2:7$. 
Then, in models where the superpotential Higgs mass $\mu\gg M_{1,2}$, one expects a
gluino of mass $m_{\tg}\sim M_3$, a wino-like chargino and 2nd lightest neutralino with mass
$m_{\tw_1,\tz_2}\sim M_2$ and a bino-like lightest neutralino with mass $m_{\tz_1}\sim M_1$.
If in addition one assumes heavy squarks (as are favored by the decoupling solution to the SUSY
flavor and $CP$ problems, the cosmological gravitino problem and proton decay), 
then for low values of $m_{\tg}\alt 1$ TeV gluino pair production is expected to be the dominant  
SUSY cross section at LHC. However, as $m_{\tg}$ increases, one samples parton distribution functions 
(PDFs) at higher values of fractional momentum $x_F$, and the gluino pair cross section drops sharply.
Meanwhile, pair production of the much lighter wino-like and bino-like states samples PDFs at much
lower $x_F$, and will suffer only a mild kinematic suppression. At some point, as $m_{\tg}$ increases, 
production of $\tw_1^+\tw_1^-$ and $\tw_1^\pm\tz_2$ will become dominant over $\tg\tg$ production.

\begin{figure}[tbp]
\postscript{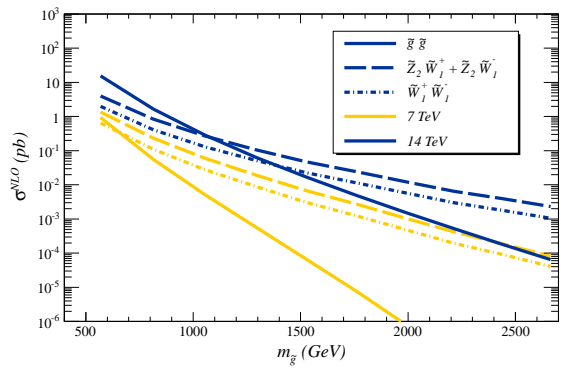}{0.95}
\caption{Total NLO cross sections for $\tg\tg$, $\tw_1^\pm\tz_2$ and $\tw_1^+\tw_1^-$
production at LHC7 (light) and LHC14 (dark), versus $m_{\tg}$, with $m_{\tq}=15$ TeV.
\label{fig:xsecs}}
\end{figure}

To illustrate, we plot in Fig.~\ref{fig:xsecs} the next-to-leading-order in QCD (NLO) 
cross sections in $pb$ 
(from Prospino\cite{prospino}) for $pp\to\tg\tg$, $\tw_1^+\tw_1^-$
and $\tw_1^\pm\tz_2$, versus $m_{\tg}$, in a SUSY model with gaugino mass unification, but with
$m_{\tq}=m_{\tell}=15$ TeV, $\tan\beta =10$ and $\mu\simeq m_{\tg}$.
The dark curves are for LHC14, while light curves are for LHC7. In this case, we see that
at LHC7, $\tw_1^\pm\tz_2$ production (dashed curves) has already become dominant for $m_{\tg}\agt 500$ GeV, 
while for LHC14, $\tw_1^\pm\tz_2$ becomes dominant for $m_{\tg}\agt 1$ TeV. As $m_{\tg}$ increases,
$\tg\tg$ production falls quickly, and gaugino pair production becomes completely dominant. 
This suggests that in the case of very heavy squark masses, 
one may want to sample the dominant cross sections, which turn out to
be gaugino pair production rather than gluino pair production.

\begin{figure}[tbp]
\postscript{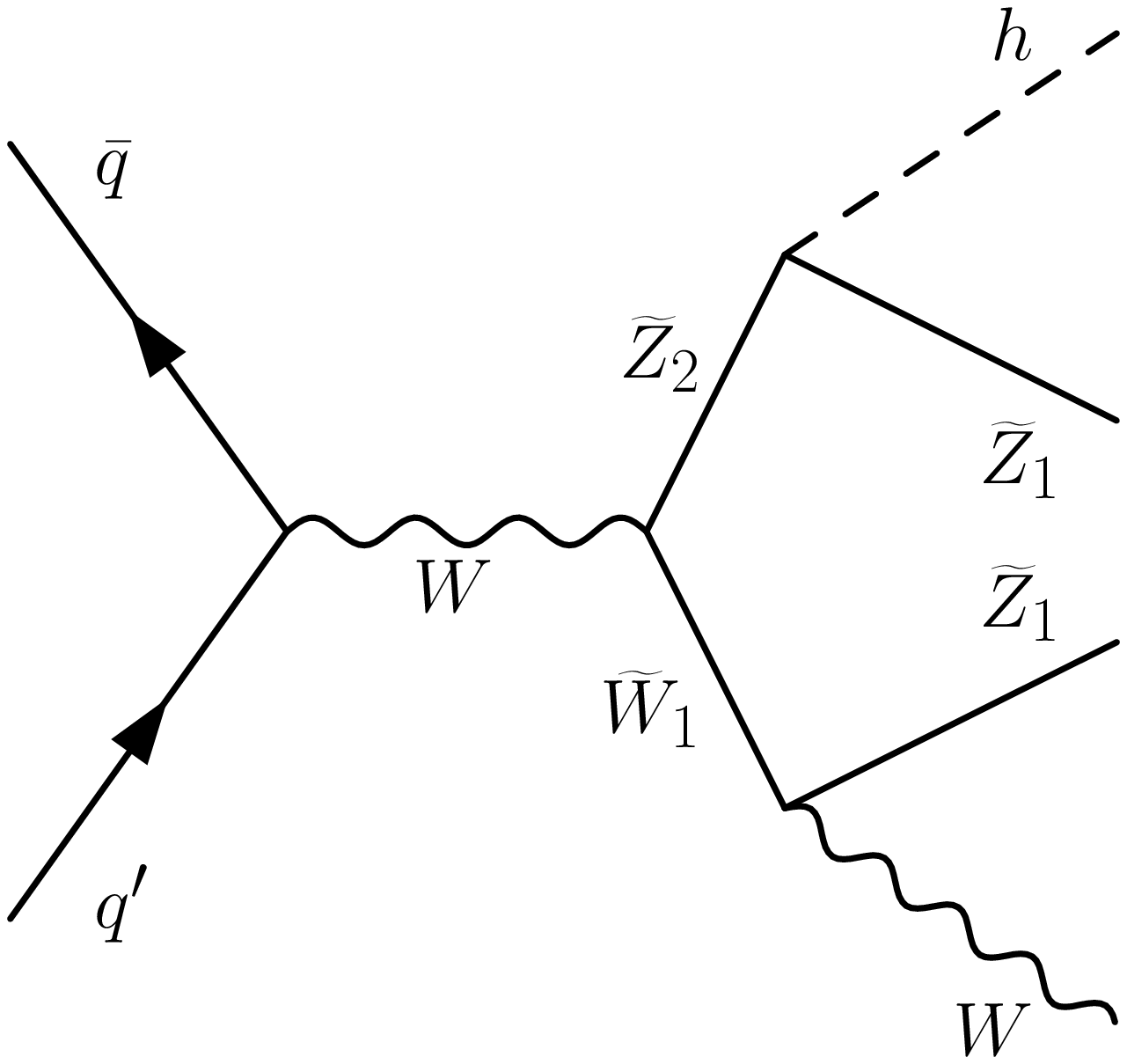}{0.6}
\caption{Feynman diagram for $q'\bar{q}\to\tw_1^\pm\tz_2\to (W^\pm\tz_1)+(h\tz_1)$
\label{fig:diag}}
\end{figure}

Now let us restrict our analysis to LHC14, for which integrated
luminosities in the 100 - 1000 fb$^{-1}$ range are expected. Assuming
models with gaugino mass unification so that $2M_1\simeq M_2$ and $\mu > M_2$, the
two-body decay $\tw_1\to W\tz_1$ with $m_{\tz_1}\sim {1\over
2}m_{\tw_1}$ is expected to dominate the $\tw_1$ branching fraction for
$m_{\tw_1}> 2M_W$, which corresponds to $m_{\tg}\agt 560$ GeV. Likewise,
the two-body decay $\tz_2\to\tz_1 h$ turns on for $m_{\tz_2}\agt
2m_h\sim 230-280$ GeV, corresponding to $m_{\tg}\agt 800-900$ GeV. The
decay $\tz_2\to\tz_1 Z$ also will occur, but usually with branching
fraction $\sim 5\%$, compared to $BF(\tz_2\to\tz_1 h )\sim 95\%$, for
the models under consideration (since $\tz_1\tz_2Z$ coupling only
involves small higgsino components of both neutralinos, whereas the
$\tz_1\tz_2h$ coupling occurs via the higgsino component of just one of
the two neutralinos).  Thus, we are led to scrutinize a single
production reaction followed by simple two-body decays:
$pp\to\tw_1^\pm\tz_2\to (W\tz_1)+(h\tz_1)\to (\ell\nu_{\ell}\tz_1)+
(b\bar{b}+\tz_1)$, as shown in Fig. \ref{fig:diag}.  Because of
potentially enormous SM backgrounds to the final state, this event topology
has never been studied previously; indeed the decay $\tz_2\to \tz_1h$
has been termed the ``spoiler mode'' in the literature.  
Here, we evaluate this signal reaction compared to SM backgrounds arising from $t\bar{t}$,
$Wb\bar{b}$, $WZ$, $Wh$ and $Zb\bar{b}$ production.
%

In our calculations, we generate sparticle mass spectra in the
mSUGRA/CMSSM model using the Isasugra\cite{isasugra} spectrum calculator
with $m_0=5$ TeV, $A_0=-1.8m_0$, $\tan\beta =10$, $\mu >0$ and with
$m_t=173.3$ GeV.  We vary $m_{\tw_1,\tz_2}$ by varying $m_{1/2}$.  We
feed the resulting IsaWIG file into the HERWIG event
generator\cite{herwig}, which maintains SUSY particle spin correlations
via preprogrammed spin density matrices\cite{richardson}.  We normalize
the signal cross section to the Prospino NLO result.  We also generate
$Wh$, $WZ$ and $t\bar{t}$ backgrounds using Herwig, and $Wb\bar{b}$,
$Zb\bar{b}$ as well as the single top\footnote{Since our signal requires
two high $E_T$ $b$ jets we have focussed on single top production from
the $q\bar{q'} \to t\bar{b}$ (or $\bar{t}b$) process with $s$-channel
$W$ exchange, and neglected contributions from $gq \to t\bar{b}q'$ and
the $gb \to tW$ processes \cite{st}.} backgrounds using an
AlpGEN\cite{alpgen}/Herwig interface.  For $t\bar{t}$ production, we use
a $k$-factor of 2 with no $k$-factors for the other backgrounds.  For
each signal and background process, we generate a statistical sample
corresponding to 100 fb$^{-1}$ of data at LHC14.

We implement the AcerDET fast detector simulation program\cite{acerdet},
using default ATLAS detector parameters including a cone-type jet
finding algorithm with $\Delta R(jet)=0.4$ and $E_T(jet)>10$ GeV.  A jet
is tagged as a $b$-jet if it contains a $b$-quark with $|\eta_b|<2.5$,
$p_T(b)>5$ GeV and the $b$ is located within $\Delta R<0.2$ around the
reconstructed jet axis.  We also impose a $b$-jet reconstruction
efficiency of 60\%, plus a $b$-jet mis-tag probability on QCD jets as in
Ref.~\cite{xt}. We then require the following pre-selection cuts ({\bf
cuts I}):
\begin{itemize}
\item exactly one isolated lepton $\ell$ ($\ell =e$ or $\mu$) with $p_T(\ell)>10$ GeV and $|\eta(\ell )|<2.5$,
\item two $b$-jets with $p_T(b-jet)>50$ GeV and $|\eta (b-jet)|<2$ (events with $\ge 3$ $b$-jets are rejected) and
\item number of non-$b$-jets with $p_T(j)>50$ GeV equals zero ($n(j)=0$).
\end{itemize}

Next, we examine a variety of distributions for a $m_{\tw_1}=620$ GeV
signal (corresponding to $m_{1/2}=700$ GeV with $m_{\tg}=1800$ GeV) and
backgrounds, including $\eslt$, $M_{eff}=\sum_{jets} E_T(jets)+\eslt$,
$\Delta\phi (b\bar{b})$ and the transverse mass $m_T(\ell,\eslt)$.  In
this case, the light Higgs mass is found to be $m_h\simeq 125$ GeV.  The
SUSY signal is expected to have a much harder $\eslt$ and $M_{eff}$
distribution than background, due to the large masses of the $\tw_1$ and
$\tz_2$ particles, and the presence of two $\tz_1$ in the final state.
In addition, since the $\tz_2$ is produced typically with
$p_T(\tz_2)\sim m_{\tz_2}$, it is expected that the $h$ from $\tz_2$
decay will be at high $p_T$, and give rise to more nearly collimated
di-$b$-jet cluster than background.  Also, the $m_T$ cut is expected to
be very effective at cutting the bulk of the background processes, since
we generally expect a Jacobian peak structure with $m_T\alt M_W$ in the
background, while the signal yields a continuum.  We find we can gain a
large background rejection while retaining much of the signal by
requiring ({\bf cuts II}):
\begin{itemize}
\item $\eslt >220$ GeV, 
\item $M_{eff}>350$ GeV,
\item $\Delta\phi (b,\bar{b})< \pi /2$ and
\item $m_T(\ell,\eslt )>125$ GeV. 
\end{itemize}

\begin{figure}[tbp]
\postscript{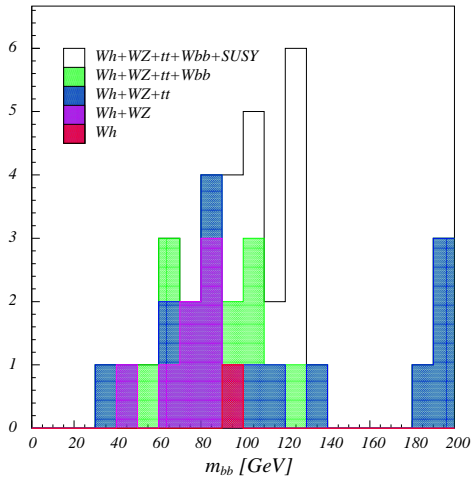}{0.95}
\caption{Number of events expected in 100 fb$^{-1}$ of LHC14 data versus $m(b\bar{b})$ for various 
summed SM backgrounds (shaded) and SUSY signal, with $m_{\tw_1}=620$ GeV and $m_h=125$ GeV. 
\label{fig:mbb}}
\end{figure}

In Fig. \ref{fig:mbb}, we plot the di-$b$-jet invariant mass
distribution after the above set of cuts I and II.  The various shaded
histograms show the $Wh$, $Wh + WZ$, $Wh + WZ + t\bar{t}$ and $Wh + WZ +
t\bar{t} + Wb\bar{b}$ backgrounds (single top and $Zb\bar{b}$ events are eliminated after
cut II).  The unshaded histogram shows the sum of all
backgrounds plus the SUSY signal for $m_{\tw_1}=620$ GeV.  From the
plot, one can see the $h\to b\bar{b}$ peak standing out beyond
background, indicating a clear signal from $\tw_1^\pm\tz_2\to
Wh\tz_1\tz_1$ production.\footnote{ Since the stabilization of the
electroweak scale prefers sub-TeV scale third generation squarks,
$bb\ell +\eslt$ events could potentially also arise from top squark pair
production although in this case the $m_{bb}$ distribution would not
peak at $m_h$.}  Both the $h$ and $Z$ peaks are located somewhat below
their naively expected positions due to jet energy loss via radiation
outside the $\Delta R=0.4$ cone, due to neutrino emission in the
$b$-decays and due to calorimeter mis-measurements.

To calculate a reach for LHC14 with 100 fb$^{-1}$, we implement an invariant 
mass cut ({\bf cut III}):
\begin{itemize}
\item $110\ {\rm GeV}< m(b\bar{b})<130\ {\rm GeV}$,
\end{itemize}
to gain a final signal sample along with background. 
A tabulation of signal and BG rates after cuts I, II and III is shown in Table \ref{tab:BG}.
We note here that the $WZ$, $Wh$ and $t\bar{t}$ backgrounds should be very well-known due to their
independent studies, and are potentially subtractable.
\begin{table}[htb]
\begin{center}
\begin{tabular}{|l|r|r|r|r|r|r|r|}
\hline
cuts &  \multicolumn{7}{c|}{ \# of events in 100 fb$^{-1}$}\\
\cline{2-8} & 
SUSY &
$t\bar t$ & $Wb\bar b$ & $WZ$& $Wh$  & $Zbb$ & total BGs\\
\hline
cuts I & 30 & 612,001 & 12,130 & 709 & 664 & 669 & 626,173\\
cuts II &  10 & 12 & 7 & 7 & 1 & 0 & 27\\ 
cuts III & 6 & 1 & 1 & 0 & 0 & 0 & 2\\ 
\hline
\end{tabular}
\end{center}
\caption{Number of events expected in 100 fb$^{-1}$ of data at LHC14 from 
SUSY signal with $m_{\tw_1}=620$ GeV and from various background processes,
after cuts I, II and III.
\label{tab:BG}}
\end{table}

\begin{figure}[tbp]
\postscript{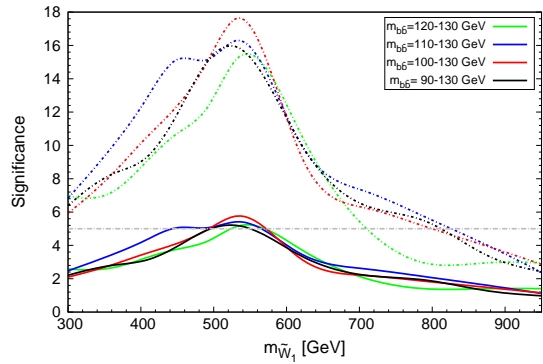}{0.95}
\caption{Significance of signal in 100 fb$^{-1}$ (solid) and 1000 fb$^{-1}$ (dashes) 
of LHC14 data versus $m_{\tw_1}$ for various 
$m(b\bar{b} )$ bin sizes. The dashed gray horizontal line shows the
$S/\sqrt{B}=5$ significance level. We have checked that whenever the
statistical significance exceeds $5\sigma$ the signal level exceeds 5
events. We take $m_0=5$ TeV and $A_0= -1.8 m_0$. 
\label{fig:sig}}
\end{figure}

The statistical significance of the signal, evaluated using Poisson
statistics, for 100 fb$^{-1}$ (solid) and 1000 fb$^{-1}$ (dashes) of
LHC14 data with several different $m(b\bar{b})$ bin sizes is shown in
Fig. \ref{fig:sig}.  Here, our signal only comes from the
$\tw_1^\pm\tz_2$ production reaction. Other SUSY production processes
would only add to these signal rates.  We see that with 100 fb$^{-1}$ of
data at LHC14, a $5\sigma$ signal emerges only for $m_{\tw_1}\sim
450-550$ GeV. However, the 1000 fb$^{-1}$ LHC14 reach extends across the
entire mass range $m_{\tw_1}\sim 300-800$ GeV.  These results require
only that weak scale gaugino masses satisfy $M_1\sim M_2/2$ and $\mu >
M_2$, since we only consider $\tw_1^{\pm}\tz_2$ production.  If we
assume the full gaugino mass unification with $M_3\sim 3.5 M_2$, then
the 100 fb$^{-1}$ range of chargino masses that is accessible at better
than the
$5\sigma$ level in Fig.~\ref{fig:sig} corresponds to $m_{\tg}\sim 1.5-1.9$
TeV, while the 1000 fb$^{-1}$ range corresponds to $m_{\tg}\sim 1-2.8$
TeV (the range of $m_{\tg}$ depends on variations within the SUSY model
parameter space).  These values turn out to be comparable to values
found in Ref. \cite{bbbkt}.  The maximal SUSY reach determined in
Ref. \cite{bbbkt} and \cite{lhc10} were found using very hard cuts, with
very low backgrounds originating from QCD processes yielding very high
jet multiplicity, for which theoretical uncertainties are quite
large. In contrast, the reach derived from $\tw_1^\pm\tz_2\to Wh+\eslt$
is determined using well-known QCD and electroweak background processes
with lower jet multiplicities for which theoretical uncertainties should
be much smaller.  In addition, since our signal involves just a single
$2\to 2$ production process followed by simple 2-body decays, the
process may allow for a $\tz_2$ mass extraction for instance from
the $p_T(h)$ distribution if a sizable event sample can be obtained.

{\it Summary:} 

For LHC running at $\sqrt{s}=14$ TeV, the dominant SUSY reaction for
$m_{\tg}\agt 1$~TeV is $pp\to\tw_1^\pm\tz_2\to Wh\tz_1\tz_1$ in models
with decoupled (heavy) scalars, gaugino mass unification and $|\mu
|>M_1,M_2$.  This reaction leads to a distinctive $\ell b\bar{b}+\eslt$
final state which can be detected above background levels for chargino
masses of 450-550~GeV, corresponding to $m_{\tg}\sim 1.5-1.9$ TeV, in
models with gaugino mass unification, for an integrated luminosity of
100 fb$^{-1}$. For a 1000 fb$^{-1}$ data sample, LHC14 should probe
chargino masses in the 300-800~GeV range corresponding to $m_{\tg} \sim
1-2.8$ TeV.  This novel signal for supersymmetry from
chargino-neutralino pair production not only serves to point toward SUSY
as the origin of any new physics that may be discovered in the canonical
multijet plus multilepton plus $\eslt$ channel, but potentially also
increases the projected SUSY reach of LHC in models where gluinos and
first generation squarks are very heavy. The simplicity of production
and decay modes begs for a $\tz_2$ mass extraction if a sufficiently
large data sample can be realized.

This work was supported in part by the U.S. Department of Energy under grant 
Nos.~DE-FG02-04ER41305, DE-FG02-04ER41291 and DE-FG02-95ER40896 and by Funda\c{c}\~{a}o de
Amparo \`{a} Pesquisa do Estado de S\~{a}o Paulo (FAPESP).



\begin{thebibliography}{99}

%
\bibitem{wss} For reviews of SUSY, see
H.~Baer and X.~Tata, {\it Weak Scale Supersymmetry: From 
Superfields to Scattering Events}, 
(Cambridge University Press, 2006); 
%
\bibitem{atlas} See {\it e.g.} G. Aad {\it et al.} (ATLAS collaboration), 
arXiv:1110.2299 (2011) and arXiv:1109.6572 (2011).
%
\bibitem{cms} See {\it e.g.} S. Chatrchyan {\it et al.} (CMS collaboration), 
arXiv:1109.2352 (2011).
%
\bibitem{msugra} 
For a review, see {\it e.g.} P. Nath, hep-ph/0307123.
%
\bibitem{bcpt1} H.~Baer, C.~H.~Chen, F.~Paige and X.~Tata, \prD{52}{1995}{2746}.
%
\bibitem{cascade} H. Baer, V. Barger, D. Karatas and X. Tata, \prD{36}{1987}{96}; 
H.~Baer, R.~M.~Barnett, M.~Drees, J.~F.~Gunion, H.~E.~Haber, D.~L.~Karatas and X.~R.~Tata,
Int.\ J.\ Mod.\ Phys.\  A {\bf 2}, 1131 (1987);
H. Baer, A. Bartl, D. Karatas, W. Majerotto and X. Tata, \ijmpa{4}{1989}{4111};
H.~Baer, X.~Tata and J.~Woodside, \prD{42}{1990}{1568};
for earlier work on sparticle decays to just gauginos, see 
H. Baer, J. Ellis, G. Gelmini, D. V. Nanopoulos and X. Tata, \plb{161}{1985}{175};
G. Gamberini, \zpc{30}{1986}{605}; H. Baer and E. Berger, \prD{34}{1986}{1361}.
%
\bibitem{bcpt2} H.~Baer, C.~H.~Chen, F.~Paige and X.~Tata, \prD{53}{1996}{6241};
H.~Baer, C.~H.~Chen, M.~Drees, F.~Paige and X.~Tata, \prD{59}{1999}{055014};
B.~Allanach, J.~Hetherington, A.~Parker and B.~Webber, 
\jhep{08}{2000}{017}.
%
\bibitem{bbbkt} H.~Baer, C.~Bal\'azs, A.~Belyaev, T.~Krupovnickas and X.~Tata,
\jhep{0306}{2003}{054}; S.~Abdullin and F.~Charles, \npb{547}{1999}{60};
S.~Abdullin {\it et al.} (CMS Collaboration), \jphg{28}{2002}{469};
%
\bibitem{lhc7} H. Baer, V. Barger, A. Lessa and X. Tata, \jhep{1006}{2010}{102}
and arXiv:1112.3044.
%
\bibitem{lhc10} H. Baer, V. Barger, A. Lessa and X. Tata, \jhep{0909}{2009}{063}.
%
\bibitem{lhc_gmsb} H. Baer, P. Mercadante, F. Paige, X. Tata and Y. Wang, 
\plb{435}{1998}{109}; H. Baer, P. Mercadante, X. Tata and Y. Wang, 
\prD{62}{2000}{095007}.
%
\bibitem{lhc_amsb} H. Baer, J. K. Mizukoshi and X. Tata,
\plb{488}{2000}{367}; A. J. Barr, C. G. Lester, A. Parker, B. Allanach 
and P. Richardson, \jhep{0303}{2003}{045}.
%
\bibitem{lhc_ino} H. Baer, A. Belyaev, T. Krupovnickas and X. Tata,
\prD{65}{2002}{075024}.
%
\bibitem{lhc_rpv} H. Baer, C. H. Chen and X. Tata, \prD{55}{1997}{1466}.
%
\bibitem{prospino}  W. Beenakker, R. Hopker, M. Spira, hep-ph/9611232 (1996).
%
\bibitem{isasugra} H. Baer, C. H. Chen, R. Munroe, F. Paige and X. Tata,
\prD{51}{1995}{1046};
F. Paige, S. Protopopescu, H. Baer and X. Tata,
hep-ph/0312045.
%
\bibitem{herwig} G. Corcella {\it et al.} (HERWIG collaboration), \jhep{0101}{2001}{010}.
%
\bibitem{richardson} P. Richardson, \jhep{0111}{2001}{029}.
%
\bibitem{st} T.~Stelzer, Z.~Sullivan and S.~Willenbrock, \prD
  {56}{1997}{5919} and \prD{58}{1998}{094021}; see also V.~Barger,
  M.~McCaskey and G.~Shaughnessy, \prD{81}{2010}{034020}.
%
\bibitem{alpgen} M. Mangano, M. Moretti, F. Piccinini, R. Pittau and
A. Polosa, \jhep{0307}{2003}{001}.
%
\bibitem{acerdet} E. Richter-Was, hep-ph/0207355 (2002).
%
\bibitem{xt} R. Kadala, P. Mercadante, J. K. Mizukoshi and X. Tata,
\epjc{56}{2008}{511}.
%
\end{thebibliography}
\end{document}